\documentclass[twocolumn,showpacs,preprintnumbers,amsmath,amssymb,floatfix]{revtex4-1}
\input{epsf}

\usepackage{graphicx}
\usepackage{dcolumn}
\usepackage{bm}

\begin{document}


\title{Quantum estimation of magnetic-field gradient using W-state}
\author{H. T. Ng and K. Kim}

\affiliation{Center for Quantum Information, Institute for Interdisciplinary Information Sciences, Tsinghua University, Beijing 100084, P. R. China}

\date{\today}


\begin{abstract}
We study the precision limits of detecting a linear 
magnetic-field gradient by using W-states in the presence of different types of noises. We consider to use 
an atomic spin chain for probing the magnetic-field gradient, 
where a W-state is prepared. We compare this method
with the measurement of using two uncorrelated atoms.
For pure states, W-states can provide an
improvement over uncorrelated states in determining 
the magnetic-field gradient up to four particles.
We examine the effects of local dephasing and dissipations 
on the performances of detections. In presence of 
dephasing, the uncorrelated 
atoms can give a higher precision than using W-states. 
But W-states provide a better
performance in the presence of dissipation for a few particles.
We briefly discuss the implementation of the detection
methods with cold atoms and trapped ions.
\end{abstract}

\maketitle

\section{Introduction}
Measurement of magnetic-field
gradient is important in magnetic
resonance imaging (MRI)  \cite{Lauterbur}
and quantum control \cite{Grinolds}, etc.
For example, MRI relies on the principle of nuclear magnetic
resonance \cite{Glover}. The information of an image
can be encoded
via the magnetic-field gradient \cite{Lauterbur}.
The quality of imaging can be improved
by reducing the signal-to-noise 
ratio of the magnetic-field gradient.
Therefore, the methods for the reduction of signal-to-noise 
ratio may advance the technology of MRI.

Precision measurements
of magnetic-field gradient \cite{Ng,Lanz,Kaler},
which use the different physical systems and 
entangled states, have recently been proposed.
In this paper, we propose to detect the 
magnetic-field gradient by using an atomic spin chain.
Indeed, neutral atoms \cite{Hinkley,Bloom} and trapped ions \cite{Rosenband} can be used to 
accurately determine the frequency 
standard which are important in science and engineering. 
Entangled states are useful to enhance the precision 
of measurements. In addition, the two 
classes of multipartite entangled states, such as
GHZ \cite{Monz} and
W-states \cite{Haffner}, have been demonstrated using trapped ions.
For example, a W-state has been produced by 
using a string of 8 trapped ions \cite{Haffner}.
More recently, a W-state of a few tens of cold atoms 
have also been produced in an optical fiber cavity \cite{Haas}.

In realistic situations, the atoms are
inevitable to be coupled to the environment.
This greatly limits the performance of
measurement. 
For example, GHZ states 
can only give the same uncertainty of
uncorrelated states in detecting
the transition frequencies
in the presence of dephasing \cite{Huelga}.  
Recently, the bounds of the measurement 
errors have been provided \cite{Escher,Dobrzanski} 
in the presence of the different types of
noises. In the limit of large number of atoms, 
the precision limit can be
improved by a constant factor in comparison with 
uncorrelated states irrespective of
the initial state and measurement scheme.
Therefore, it is important to study the precision limits 
of the different input states in the presence of
the different types of noises.

In this paper, we propose a method to detect
the magnetic-field gradient
by using a chain of atoms, where the
atoms are prepared in a W-state. 
W-states are genuine multipartite entangled 
states \cite{Dur} and are robust against noises 
and particle losses \cite{Haffner}. 
We show that the magnetic-field gradient
can be encoded onto the coherence of
an atomic spin chain. Therefore, it can be used for 
measuring the magnetic-field gradient.
Also, we compare this method with another detection method by
using two uncorrelated atoms.

In fact, the precision limits of detecting 
the magnetic-field gradient by using uncorrelated states 
and W-states are both inversely proportional to the size
of the system. To compare the two methods
with the same size,
the distance between two separate uncorrelated
atoms is equal to the length of a chain of atoms
which are prepared in a W-state.
We show that W-states can provide an improvement 
over the method
by using uncorrelated states up to four particles.

We investigate the effects of local dephasing
and dissipations on these two detection methods.
In contrast to the cases of pure states, 
their performances are different in the presence
of the different types of noises.
We find that uncorrelated states are able
to give better performance than W-states 
when the dephasing noise is present.
However, W-states provide a higher precision than
uncorrelated states for a few particles 
in the presence of dissipation.

This paper is organized as follows: 
In section 2, we introduce the system of
an atomic spin chain and discuss the coupling of
atoms to a linear gradient magnetic-field. 
In section 3, we study the precision limits
of using an atomic spin chain for measuring
the magnetic-field gradient, where the atoms
are in prepared in a W-state. We compare
this method with the measurement by using
two uncorrelated atoms.
In section 4, we investigate the effects of 
local dephasing and dissipation 
on the performances of the two different 
detection methods.
In section 5, we briefly discuss the physical realization
of a spin chain by using cold atoms and trapped ions. 
We provide a summary in section 6. 
We give out the details of calculations in appendices.

\section{Detection of magnetic-field gradient}
\subsection{System}
\begin{figure}[ht]
\centering
\includegraphics[height=3.0cm]{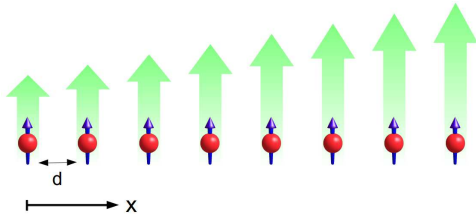}
\caption{ \label{spin_chain} (Color online) 
An atomic spin chain is coupled to a linear magnetic-field 
gradient in the $x$-direction. Each atom is separated with a distance $d$.}
\end{figure}

We consider an atom to have two hyperfine spin states.
When an atom is coupled to a magnetic field,
the energy splitting between two hyperfine
states is changed due to the Zeeman effect \cite{Glover}.
By measuring the difference of the transition 
frequencies of two atoms at the two different locations,
the magnetic-field gradient can be determined. 
In general, an atomic spin 
chain can be used for detecting the 
magnetic-field gradient
as shown in Fig.~\ref{spin_chain}, where
each atom is equally spaced with a distance $d$.
We assume that the magnetic field $B(x_j)$ 
linearly varies with the position $x_j$ as
\begin{eqnarray}
B(x_j)&=&B_1+Gx_j,
\end{eqnarray}
where $B_1$ is the reference magnetic field
and $G$ is the magnetic-field gradient. 
The Hamiltonian, which describes the coupling
between atom and the magnetic field, is 
\begin{eqnarray}
\label{Ham}
H&=&\hbar\sum^N_{j=1}\omega_{j}\sigma^j_z,
\end{eqnarray}
where $\omega_j$ and $\sigma^j_z$ are 
the energy frequency and Pauli operator 
of atom $j$.  The total number of atoms in
the chain is $N$. The transition frequency 
$\omega_j$ depends on the magnitude of 
the magnetic field $B(x_j)$ as
\begin{eqnarray}
\omega_j=\omega_0+\gamma{B(x_j)},
\end{eqnarray} 
where $\omega_0$ is the transition frequency of 
the two hyperfine spin states without the external
magnetic-field and $\gamma$ is the gyromagnetic ratio of
an atom.

\subsection{Using W-states}
We study the detection method by using
W-states. Initially, the system is prepared 
in a W-state which is written as
\begin{eqnarray}
\label{w_j}
|W\rangle&=&\frac{1}{\sqrt{N}}(|100\ldots{0}\rangle+|010\ldots{0}\rangle+\ldots+|000\ldots{1}\rangle),\nonumber\\
&=&\frac{1}{\sqrt{N}}\sum^N_{j=1}|w_j\rangle,
\end{eqnarray}
where $|w_j\rangle=|1\rangle_j\prod_{j'}|0\rangle_{j'}$ 
and $j'\neq{j}$.
When the atomic chain is coupled to the 
magnetic-field gradient for a time $t$, the state becomes
\begin{eqnarray}
|\psi(t)\rangle&=&\frac{1}{\sqrt{N}}\sum^N_{j=1}e^{-2ij\theta_1{t}}|w_j\rangle,
\end{eqnarray}
where $\theta_1=G\gamma{d}$. The global
phase factor has been omitted here.

The magnetic-field gradient can be 
encoded onto the quantum coherence
of the state. The quantum
coherence factor $C_1$ can be defined as
\begin{eqnarray}
\label{Coherence_est}
C_1&=&N|W\rangle\langle{W}|-\sum^N_{j=1}|w_j\rangle\langle{w}_j|.
\end{eqnarray}
Now $\langle{C_1}\rangle$ can be expressed as:
\begin{eqnarray}
\label{meanC}
\langle{C_1}\rangle&=&-1+\frac{1}{N}\frac{\sin^2(N\theta_1{t})}{\sin^2(\theta_1{t})}.
\end{eqnarray}
The quantity $\langle{C_1}\rangle$ is a function of the parameter 
$\theta_1$ which is proportional to the gradient $G$.
Therefore, the magnetic-field gradient can be determined
from the coherence $\langle{C_1}\rangle$.
The details of the derivation of Eq.~(\ref{meanC}) 
is provided in Appendix A. In fact, the quantity $\langle{C_1}\rangle$
can be obtained by averaging the terms 
$\langle{\prod_{k\neq{i,j}}|0\rangle_{k}{}_{k}\langle{0}|\rangle\langle\sigma^i_-\sigma^j_++\sigma^j_-\sigma^i_+}\rangle$
which can be experimentally measured.

\begin{figure}[ht]
\centering
\includegraphics[height=5.0cm]{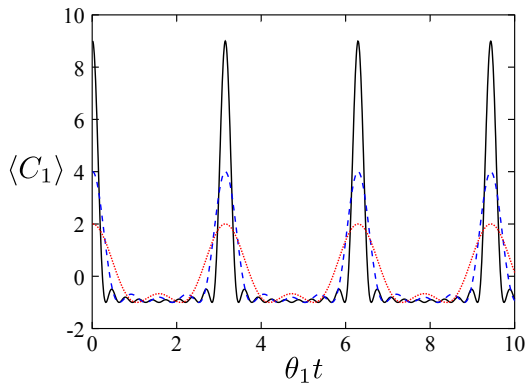}
\caption{ \label{fig_C1} (Color online) $\langle{C_1}\rangle$ versus 
time $\theta_1{t}$, for the different number of atoms $N$.
The different number of atoms $N$ are denoted by
the different lines: $N=3$ (red-dotted), $N=5$ (blue-dashed)
and $N=10$ (red-solid), respectively.}
\end{figure}

In Fig.~\ref{fig_C1}, we plot the coherence $\langle{C_1}\rangle$ versus time, for
the different number of atoms $N$. The small 
oscillations are observed when $\langle{C_1}\rangle$ is about -1. 
The peaks occur when the dimensionless time 
$\theta_1{t}$ is a multiple of $\pi$. The 
maximum of peaks can attain $N-1$.

The magnetic-field gradient can be determined from the coherence 
$\langle{C_1}\rangle$.
The uncertainty $\delta\Theta_1$ of the measurement is
\begin{eqnarray}
\delta\Theta_1&=&\dfrac{\Delta{C_1}}{\bigg|\dfrac{\partial\langle{C_1}\rangle}{\partial\Theta_1}\bigg|},\\
\label{unphi}
&=&\frac{[N^2\sin^2\Theta_1-\sin^2(N\Theta_1)]^{1/2}|\sin(N\Theta_1)|}{|N\sin(2N\Theta_1)-2\cot\Theta_1\sin^2(N\Theta_1)|}.
\end{eqnarray}
where $\Theta_1=\theta_1{t}$ and $\Delta{C_1}=\sqrt{\langle{C^2_1}\rangle-\langle{C_1}\rangle^2}$.

The minimum of
uncertainty $\delta\Theta_{\rm min}$ is obtained
\begin{eqnarray}
\label{dTheta}
\delta\Theta^{\rm min}_1&=&\frac{1}{2}\sqrt{\frac{3}{N^2-1}},
\end{eqnarray}
when $\Theta_1={n}\pi$ and $n$ is an integer.
The minimum uncertainty is proportional to 
$N^{-1}$ for large $N$. The minimum uncertainty 
in Eq.~(\ref{dTheta})
can be intuitively understood from the time-energy
uncertainty relation \cite{Giovannetti0}.
The minimum uncertainty $\delta\Theta_{\rm min}$
occurs when the energy fluctuation of the state 
is maximum.
The bound in Eq.~(\ref{dTheta}) can be derived
from the time-energy uncertainty which is given 
in Appendix B.

\subsection{Using two uncorrelated atoms}
\begin{figure}[ht]
\centering
\includegraphics[height=5.0cm]{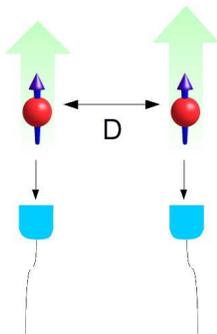}
\caption{ \label{uncorrelated_ions} (Color online) 
Two uncorrelated atoms are separate with a distance
$D$ for measuring the magnetic-field gradient. The two
different magnetic fields are independently measured.}
\end{figure}

We study the measurement of magnetic-field gradient 
by using two uncorrelated atoms which are separated
with a distance $D$ as shown in Fig.~\ref{uncorrelated_ions}. 
We will compare the precision limits of using
two uncorrelated atoms and an atomic chain with the
same system's size $D$. To facilitate our
subsequent discussion, we set $D=(N-1)d$, where
$N$ is the number of atoms in the chain.

The magnetic-field gradient can be determined
by individually measuring two atoms at the two 
different positions. The magnetic-field gradient is
\begin{eqnarray}
G=\frac{B_2-B_1}{D},
\end{eqnarray}
where $B_2$ and $B_1$ are the magnetic
fields at the two different positions.

Now, we briefly describe the procedures
of measurement. Initially, the atoms
are prepared in the equally weighted
superpositions as
\begin{eqnarray}
|\Psi(0)\rangle_j&=&\frac{1}{\sqrt{2}}(|0\rangle_j+|1\rangle_j),
\end{eqnarray} 
where $j=1$ and 2.
Then, let the system freely evolve for 
a time $t$. The state becomes
\begin{eqnarray}
|\Psi(t)\rangle_j&=&\frac{1}{\sqrt{2}}(e^{i\omega_j{t}}|0\rangle_j+e^{-i\omega_j{t}}|1\rangle_j).
\end{eqnarray} 
By applying a $\pi/2$-pulse to the atom, the state can be written as
\begin{eqnarray}
|\Psi'(t)\rangle_j&=&\frac{1}{2}e^{i\omega_j{t}}[(1-e^{2i\omega_j{t}})|0\rangle+(1+e^{-2i\omega_j{t}})|1\rangle].\nonumber\\
\end{eqnarray} 
The frequency $\omega_j$ can be determined by measuring the
probability $P_j$ of the state $|1\rangle_j$ which is given by
\begin{equation}
P_j=\frac{1}{2}(1+\cos2\omega_j{t}).
\end{equation}

The uncertainty $\delta\omega_j$ is given by \cite{Huelga}
\begin{eqnarray}
\delta\omega_j&=&\sqrt{{P_j(1-P_j)}}\Bigg|\frac{dP}{d\omega_j}\Bigg|^{-1},\\
&=&\frac{1}{2t}\Bigg[\frac{1-\cos^2(2\omega_j{t})}{\sin^2(2\omega_j{t})}\Bigg]^{1/2}.
\end{eqnarray}
When $t={m}\pi/\omega_j'$, the minimum uncertainty can be obtained,
where $m$ is an odd integer. The minimum uncertainty $\delta\omega^{\rm min}_j$ is
\begin{eqnarray}
\label{pdomega}
\delta\omega^{\rm min}_j=\frac{1}{2t}.
\end{eqnarray}

The uncertainty of the gradient is proportional
to the sum of $\delta\omega_1$ and $\delta\omega_2$,
\begin{eqnarray}
\delta{G}=\frac{\sqrt{\delta{\omega^2_2}+\delta{\omega^2_1}}}{\gamma{D}},
\end{eqnarray}
where $\delta\omega_j=\gamma\delta{B_j}$.
We can express the uncertainty $\delta\theta_2$ as
\begin{eqnarray}
\label{sumdtheta}
\delta\theta_2&=&\frac{\sqrt{\delta\omega^2_2+\delta\omega^2_1}}{N-1},
\end{eqnarray}
where $\theta_2=\gamma{G}d$ and $D=(N-1)d$. From
Eqs.~(\ref{pdomega}) and (\ref{sumdtheta}),
the minimum uncertainty $\delta\Theta^{\rm min}_2$ is
\begin{eqnarray}
\label{local}
\delta\Theta^{\rm min}_2&=&\frac{1}{\sqrt{2}(N-1)},
\end{eqnarray}
where $\Theta_2=\theta_2{t}$.

\subsection{Comparsion of two methods}
\begin{figure}[ht]
\centering
\includegraphics[height=5.0cm]{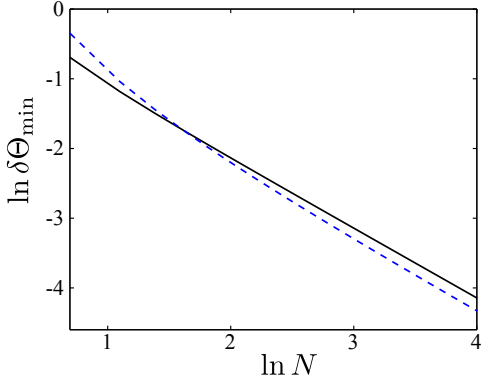}
\caption{ \label{puncert} (Color online) Log-log plot of $\delta\Theta_{\rm min}$
versus $N$. The W-states and uncorrelated states are denoted with black solid and blue dashed lines, respectively.}
\end{figure}

Now we compare the performances of these
two methods.
In Fig.~\ref{puncert}, we plot the minimum uncertainties
of uncorrelated states with the two different methods.
Although the uncertainties of the
two methods both scale with $N^{-1}$ if $N$ is large.
W-states can provide a higher precision than using
uncorrelated atoms up to 4 particles.

In addition, it should be noted that the improving factor
$N^{-1}$ comes from the gradient field which gives
rise to the energy fluctuation of the state of order
of $N$. 
The uncorrelated
states can also provide the scaling 
$N^{-1}$. 
Therefore, in this sense, the scaling of $N^{-1}$ is 
different to the usual Heisenberg-limited measurement
\cite{Giovannetti}.

\section{Effects of dephasing and dissipation}
Now we study the effects of local 
dephasing and dissipation of atoms 
on measurements. 
We consider each atom to be independently
coupled to the bath. We have assumed that the 
Markovian approximation is valid. 
The master equation of atom $j$, which describes
dissipation and dephasing at low temperature, is written as \cite{Barnett}
\begin{eqnarray}
\dot{\rho}^{(j)}&=&i[\rho^{(j)},H]+{\Gamma_p}(\sigma^j_z\rho^{(j)}\sigma^j_z-\rho)+\frac{\Gamma_d}{2}(2\sigma^j_-\rho^{(j)}\sigma^j_+\nonumber\\
&&-\sigma^j_+\sigma^j_-\rho^{(j)}-\rho^{(j)}\sigma^j_+\sigma^j_-),
\end{eqnarray} 
where $\Gamma_p$ and $\Gamma_d$ are the dephasing and dissipation rates, respectively.
The dissipative dynamics can be solved as:
${\rho}^{(j)}_{11}=e^{-\Gamma_d{t}}\rho^{(j)}_{11}(0)$,
${\rho}^{(j)}_{00}=1-e^{-\Gamma_d{t}}\rho^{(j)}_{11}(0)$ and
${\rho}^{(j)}_{01}=e^{-\frac{\Gamma_d}{2}{t}-\Gamma_p{t}+2i\omega_j{t}}\rho^{(j)}_{01}(0)$.

The system becomes incoherent and the dynamics
is no longer periodic. Therefore,
the precision limits are limited due to 
noises. 
For W-states, we can obtain the analytical expression 
of the uncertainty of the parameter $\theta_1$
in the presence of dissipation and dephasing,
It is given by
\begin{equation}
\label{theta_decoh}
\delta\theta_1\!\!=\!\!\frac{[g_1(t)\sin^4\!{\Theta_1}\!+\!g_2(t)\!\sin^2\!\Theta_1\sin^2(\!N\Theta_1\!)\!-\!\sin^4(\!N\Theta_1\!)]^{\frac{1}{2}}}{t|N\sin(2N\Theta_1)-2\cot\Theta_1\sin^2(N\Theta_1)|},
\end{equation}
where 
$g_1(t)=N^2[(N-1)e^{(2\Gamma-\Gamma_d){t}}-(N-2)e^{\Gamma{t}}-1]$,
$g_2(t)=[N(N-2)+2Ne^{-\Gamma{t}}]e^{\Gamma{t}}$ and $\Gamma=\Gamma_d+2\Gamma_p$.

For uncorrelated atoms,
the uncertainty is given by  
\begin{equation}
\label{dindividual}
\delta\theta_2=\frac{1}{t(N-1)}\Bigg[\sum^2_{j=1}\frac{1-e^{-\Gamma{t}}\cos^2(2\omega_j{t})}{4e^{-\Gamma{t}}\sin^2(2\omega_j{t})}\Bigg]^{1/2},
\end{equation}
where $\Gamma=\Gamma_d+2\Gamma_p$.
From Eq.~(\ref{dindividual}), the minimum uncertainty occurs when $t=\Gamma^{-1}$ and 
$\omega_j{t}=n\pi/2$, where $n$ is odd. It is given by
\begin{eqnarray}
\delta\theta^{\rm min}_2&=&\frac{1}{N-1}\sqrt{\frac{{\Gamma^2{e}}}{2}}.
\end{eqnarray}

\begin{figure}[ht]
\centering
\includegraphics[height=9.0cm]{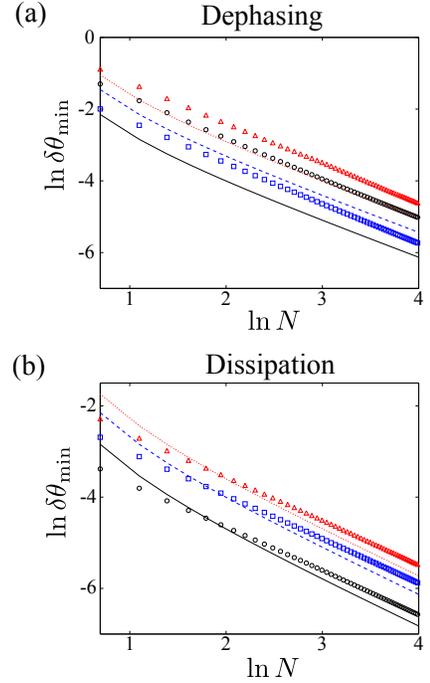}
\caption{ \label{dephase_dissipationfig1} (Color online) Log-log 
plot of $\delta\theta_{\rm min}$ versus ${N}$. The dimensionless
time $\theta{t}$ is used. The cases of dephasing and dissipation are shown in (a) and (b), respectively. 
The different dephasing ($\Gamma_p$) and dissipation ($\Gamma_d$) rates, using W-states, are denoted by
the different symbols: $0.05\theta$ (black circle), $0.1\theta$ (blue square) and
$0.15\theta$ (red triangle), respectively. 
The measurements, using uncorrelated atoms, are denoted by
the different lines:  $0.05\theta$ (black solid), $0.1\theta$ (blue dashed) and
$0.15\theta$ (red dotted), respectively.
}
\end{figure}

We numerically search the local minimum
of the uncertainties in Eq.~(\ref{theta_decoh}) within
the dimensionless time interval, $\theta{\Delta{t}}$, between 0 and $\pi$,
where the smallest time step is up to $10^{-7}$.
There exists a minimum point around $\theta_1{t}=\pi$
but the time interval is shorter than $10^{-7}$.
We have abandoned those points which are unstable (the first 
derivative of those points vary rapidly instead of
being zeros). 

We compare the performance of using uncorrelated
states and W-states in 
Fig.~\ref{dephase_dissipationfig1}.
We plot $\delta\theta_{\rm min}$ versus ${N}$ on a
logarithmic scale, for the different dephasing 
and dissipation rates in Figs.~\ref{dephase_dissipationfig1}
(a) and (b), respectively.  
If the dephasing and dissipation rates are much smaller
than the parameter of magnetic-field gradient, 
than $\ln\delta\phi_{\rm min}$ of
W-states is proportional to $\ln{N}$ and the slope is about -1. 
The uncertainty is proportional to $N^{-1}$. 

For pure states, W-states can provide a higher precision than uncorrelated states up to 4 particles.  
In Fig.~\ref{dephase_dissipationfig1}(a), the uncorrelated states can do better than W-states in the presence of dephasing. 
In Fig.~\ref{dephase_dissipationfig1}(b), the precision
limits of W-states and uncorrelated states in the presence
of dissipations are plotted. W-states can provide a higher
precision than uncorrelated states for a few particles. 
This result shows that the different
precision limits of W-states and uncorrelated states
can be obtained in the presence of the different
types of noises.

\section{Discussion}
Let us briefly discuss the methods of
realizations of a spin chain using cold atoms
and trapped ions, respectively.
An atomic spin chain can be realized
by using ultracold atoms in an optical lattice \cite{Fukuoka}.
Indeed, an atomic spin chain has recently been
shown by using a bosonic Mott insulator.
In principle, an order of hundred of atoms
can be confined in a 1D lattice. 
However, this may be difficult
to load exactly one atom into
every lattice site.
To produce a W-state in an optical lattice, nonlocal
interactions between the atoms are
necessary. This can be made by
coupling the atoms to a cavity mode \cite{Haas}.
In fact, the atoms, strongly coupling
to a cavity mode, have been demonstrated
in experiments \cite{Haas,Brennecke,Colombe}. 
Also, the W-states can be generated in
a spin chain with ferromagnetic interaction \cite{Brus}.
The specific spin-spin interactions can be
engineered in optical lattices. 

Alternatively, a spin chain can be
implemented by using trapped ions.
In fact, a W-state has been created in
a string of trapped ions \cite{Haffner} which 
can be used for detecting the magnetic-field
gradient. An order of several tens of ions \cite{Islam}
can be realized in the near future. 
However, the distance 
between ions are not equally spaced. Also,
micromotions of ions may introduce
an additional noise to the measurements. 
This may be resolved by either using
an anharmonic linear trap \cite{Lin} or 
combining an additional 
optical lattice \cite{Linnet} 
to adjust the position of ions.

Next, we discuss the measurements of estimators by using 
the two different methods.
In the uncorrelated case, it is necessary to perform two independent measurements at the two ends for the gradient-field measurement. By using a W-state, the magnetic-field gradient can be determined by just measuring the coherence factor. To  facilitate our discussion, we write a collection of qubits in terms of angular momentum operators $(J_x,J_y,J_z)$. It is easy to show that the coherence factor $C_1$ can be related to $J^2_x$ plus a constant, where $J_x=\sum_i\sigma^i_x$. To probe the variance $J^2_x$, all qubits can be coupled to a polarized light. The correlation information can be mapped onto the output field. Thus, the variance $J^2_x$ can be determined by a collective quantum-nondemolition measurement. This detection method has been proposed \cite{Eckert}. In this way, the effort of measurement on the qubits can be much reduced. Bearing in mind that the generation of W-states in a cavity can benefit from collective measurements in \cite{Haas}. This method may offer the advantages by using advanced experimental techniques in quantum optics such as collective quantum-nondemolition measurements \cite{Eckert} and cavity enhanced interactions \cite{Haas}.

In addition, this scheme can be generalized to detect the nonlinear gradient field. For example, the magnetic field 
$B(x)$ is proportional to $x^\alpha$, where $x$ is the position and $\alpha$ is the power. The uncertainty of magnetic-field gradient is proportional to $1/N^\alpha$. It is because the uncertainty of the parameter is proportional to the energy fluctuation 
of the gradient field. This is similar to the discussion for a linear field in the Appendix.

\vspace{20 mm}

\section{Conclusion}
In summary, we have studied and compared
the two different methods on measurement 
of magnetic-field gradient by using cold atoms.
We investigate an atomic chain, which is prepared
in a W-state, for detection the magnetic-field gradient. 
We also study the detection method by using two uncorrelated 
atoms. For pure states, W-states can provide
a higher precision than uncorrelated states.
In addition, we have studied 
the effect of local dephasing and dissipations 
on the performances of the two different detection methods. 
In contrast to the cases of pure states, we find 
that uncorrelated atoms can 
provide a better performance than using W-states 
in the presence of dephasing. 
But W-states can provide a higher precision
than uncorrelated states for a few particles 
if the dissipation is present.
This result shows that the precision
limits are greatly changed due to the different types
of noises.
Finally, we have briefly discussed 
the methods of realizations using cold atoms and 
trapped ions.

\section*{Acknowledgments}
We thank for I\~{n}igo Urizar-Lanz, 
Iagoba Apellaniz, and Geza Toth for their
useful comments.
This work was supported in part by the 
National Basic Research Program of 
China Grant 2011CBA00300, 2011CBA00301, 2011CBA00302
the National Natural Science Foundation of 
China Grant 11304178, 61073174, 61033001, 61061130540,
61361136003.
KK acknowledges the support from young 1000 plan.


\appendix

\section{Derivation of expectation value and variance of the coherence}
We provide a closed form for the summation of cosine functions \cite{wolfram}.
We write
\begin{equation}
\sum^{N-1}_{j'=1}\sum^N_{j=j'+1}\cos[2({j'}-{j})\theta_1{t}]
=\sum^{N-1}_{j'=1}\sum^{j'}_{j=1}\cos(2j\theta_1{t})
\end{equation}
From the Dirichlet kernel, we have
\begin{eqnarray}
1+2\sum^{j'}_{j=1}\cos(2j\theta_1{t})&=&\frac{\sin[(2j'+1)\theta_1{t}]}{\sin(\theta_1{t})},\\
\sum^{j'}_{j=1}\cos(2j\theta_1{t})&=&\frac{1}{2}\Bigg[\frac{\sin[(2j'+1)\theta_1{t}]}{\sin(\theta_1{t})}-1\Bigg].
\end{eqnarray}
Thus, 
\begin{eqnarray}
\label{cos_sum}
&&\sum^{N-1}_{j'=1}\sum^{j'}_{j=1}\cos(2j\theta_1{t})\nonumber\\
&=&-\frac{N-1}{2}+\frac{1}{2\sin(\theta_1{t})}\sum^{N-1}_{j'=1}\sin[(2j'+1)\theta_1{t}].
\end{eqnarray}
Also, from the identity, 
\begin{eqnarray}
\sum^{N}_{j'=1}\sin[(2j'-1)\theta_1{t}]&=&\csc(\theta_1{t})
\sin^2(N\theta_1{t}),
\end{eqnarray}
and therefore
\begin{eqnarray}
&&\sum^{N-1}_{j'=1}\sin[(2j'+1)\theta_1{t}]\nonumber\\
&=&\sum^{N}_{j'=1}\sin[(2j'-1)\theta_1{t}]-\sin(\theta_1{t}),\\
\label{sinId}
&=&\csc(\theta_1{t})\sin^2(N\theta_1{t})-\sin(\theta_1{t}).
\end{eqnarray}
Substitute Eq.~(\ref{sinId}) into Eq.~(\ref{cos_sum}), we obtain
\begin{eqnarray}
\label{sumcosine}
\sum^{N-1}_{j'=1}\sum^{j'}_{j=1}\cos(2j\theta_1{t})&=&-\frac{N}{2}+\frac{\sin^2(N\theta_1{t})}{2\sin^2(\theta_1{t})}.
\end{eqnarray}
From Eq.~(\ref{sumcosine}), $\langle{C_1}\rangle$ is given by
\begin{eqnarray}
\langle{C_1}\rangle&=&\frac{2}{N}\sum^{N-1}_{j=1}\sum^N_{j'=j+1}\cos[2({j'}-{j})\theta_1{t}],\\
&=&-1+\frac{1}{N}\frac{\sin^2(N\theta_1{t})}{\sin^2(\theta_1{t})},
\end{eqnarray}
and $\langle{C^2_1}\rangle$ is 
\begin{eqnarray}
&&\langle{C^2_1}\rangle\nonumber\\
&=&N-1+\frac{2(N-2)}{N}\sum^{N-1}_{j'=1}\sum^{j'}_{j=1}\cos(2j\theta_1{t}),\\
&=&N-1+\frac{2(N-2)}{N}\Big[-\frac{N}{2}+\frac{\sin^2(N\theta_1{t})}{2\sin^2(\theta_1{t})}\Big],\\
&=&1+\frac{(N-2)}{N}\frac{\sin^2(N\theta_1{t})}{\sin^2(\theta_1{t})}.
\end{eqnarray}

The derivative of $\langle{C_1}\rangle$, with respect to $\theta_1$, is
\begin{eqnarray}
&&\frac{\partial}{\partial\theta_1}\langle{C_1}\rangle\nonumber\\
&=&\frac{t}{N\sin^2(\theta_1{t})}[N\sin{2N\theta_1{t}}-2\cot(\theta_1{t})\sin^2(N\theta_1{t})].\nonumber\\
\end{eqnarray} 

\section{Bound of uncertainty from time-energy uncertainty relation}
We can obtain the bound of uncertainty $\delta\Theta$ from
the time-energy uncertainty relation \cite{Giovannetti0}:
\begin{eqnarray}
\delta\varphi\delta{E}\leq\frac{1}{2\sqrt{M}},
\end{eqnarray}
where $\delta\varphi$ and $\delta{E}$ are
the uncertainties of the time $\varphi$ and
energy of the input state, respectively, and
$M$ is the number of times for repeating the
experiments.

We calculate the variance of the energy
Hamiltonian in Eq.~(\ref{Ham}). The
expectation values $\langle{H}\rangle$
and $\langle{H^2}\rangle$ are given by
\begin{eqnarray}
\langle{H}\rangle&=&\langle{W}|{H}|{W}\rangle,\\
\label{H}
&=&\frac{\hbar}{N}(2-N)\sum_j\omega_j,\\
\langle{H^2}\rangle&=&\langle{W}|{H^2}|{W}\rangle,\\
\label{H2}
&&\frac{\hbar^2}{N}\Big[4\sum_j\omega^2_j+(N-4)\Big(\sum_j\omega_j\Big)^2\Big],
\end{eqnarray}
where $\omega_j=\omega_0+\gamma(j-1)d$.
The summations of $\omega_j$ and $\omega^2_j$ are
\begin{eqnarray}
\label{omega}
\sum_j\omega_j&=&\omega_0{N}+\frac{\gamma{d}}{2}N(N-1),\\
\label{omega2}
\sum_j\omega^2_j&=&\omega^2_0{N}+\gamma{d}\omega_0N(N-1)\nonumber\\
&&+(\gamma{d})^2\Big[\frac{1}{6}N(N+1)(2N-5)+N\Big].
\end{eqnarray} 
By using Eqs~(\ref{H})-(\ref{omega2}), the variance
$\delta{E}=\Delta{H}$ is given by
\begin{eqnarray}
\label{DH2}
\Delta{H}&=&{(\hbar\gamma{d})}\sqrt{\frac{N^2-1}{3}}.
\end{eqnarray}

From Eq.~(\ref{DH2}), we can obtain the bound of $\delta{\Theta}=\gamma{d}\delta\varphi/\hbar$ as
\begin{eqnarray}
\delta\Theta_{\rm min}&=&\frac{\hbar}{2\Delta{H}},\\
\label{TETheta}
&=&\frac{1}{2}\sqrt{\frac{3}{N^2-1}}.
\end{eqnarray} 
We find that this expression in Eq.~(\ref{TETheta}) 
coincides the bound in Eq.~(\ref{dTheta}). 
This means that the estimator $C_1$ in Eq.~(\ref{Coherence_est}) can 
give out the best precision.

\section{Expectation values and variances in the presence of dephasing and dissipation}
The expression of $\langle{C_1}\rangle$,
$\langle{C^2_1}\rangle$ and $\frac{\partial}{\partial\theta_1}\langle{C_1}\rangle$,
in the presence of dephasing and dissipation, are given by
\begin{eqnarray}
&&\langle{C_1}\rangle\nonumber\\
&=&\frac{2}{N}\sum^{N-1}_{j=1}\sum^N_{j'=j+1}\cos[2({j'}-{j})\theta_1{t}]e^{-\Gamma{t}},\\
&=&\Big[-1+\frac{1}{N}\frac{\sin^2(N\theta_1{t})}{\sin^2(\theta_1{t})}\Big]e^{-\Gamma{t}}.\\
&&\langle{C^2_1}\rangle\nonumber\\
&=&\frac{2(N-2)}{N}\sum^{N-1}_{j=1}\sum^N_{j'=j+1}\cos[2({j'}-{j})\theta_1{t}]e^{-\Gamma{t}}\nonumber\\
&&+(N-1)e^{-\Gamma_d{t}},\\
&=&(N-1)e^{-\Gamma_d{t}}+\Big[1+\frac{N-2}{N}\frac{\sin^2(N\theta_1{t})}{\sin^2(\theta_1{t})}\Big]e^{-\Gamma{t}},\\
&&\frac{\partial}{\partial\theta_1}\langle{C_1}\rangle\nonumber\\
&=&\frac{te^{-\Gamma{t}}}{N\sin^2(\theta_1{t})}[N\sin{(2N\theta_1{t})}-2\cot(\theta_1{t})\sin^2(N\theta_1{t})],\nonumber\\
\end{eqnarray}
where $\Gamma=\Gamma_d+2\Gamma_p$.

\end{document}